\documentclass[12pt]{article}

\RequirePackage{color}
\definecolor{MyDarkGreen}{rgb}{0.02,0.60,0.06}

\title{\bf Extended Scaling in High Dimensions}
\author{ 
{\it B.~Berche$^{\,1}$,} {\it C.~Chatelain$^{\,1}$,} {\it C.~Dhall$^{\,2}$,}\\
{\it R.~Kenna$^{\,1,2}$,}  {\it R.~Low$^{\,2}$,} and {\it J.-C.~Walter$^{\,1}$}\\~\\
$^1$ Laboratoire de Physique des Mat{\'{e}}riaux, UMR CNRS No. 7556, \\
Universit{\'{e}} Henri Poincar{\'{e}} (Nancy 1), B.P. 239,\\ 
F-54506 Vand{\oe}uvre l{\`{e}}s Nancy cedex, France
{}\\~\\
$^2$ Applied Mathematics Research Centre,
Coventry University,\\
Coventry, CV1 5FB, England
{}\\~\\
 }
\begin{document}
\maketitle
                      {\Large
                      \begin{abstract}
%
We apply and test the recently proposed ``extended scaling'' scheme
in an analysis of the magnetic susceptibility of Ising systems above
the upper critical dimension. The data are obtained by Monte Carlo
simulations using both the conventional Wolff cluster algorithm and the
Prokof'ev-Svistunov worm algorithm. As already observed for other models,
extended scaling is shown to extend the high-temperature critical scaling
regime over a range of temperatures much wider than that achieved
conventionally. It allows for an accurate determination of leading and
sub-leading scaling indices, critical temperatures and amplitudes of the
confluent corrections.
%
                        \end{abstract} }
%
\thispagestyle{empty}
%
%
  \newpage
%
                  \pagenumbering{arabic}

\section{Introduction}
\label{es}
\setcounter{equation}{0}

An approach to critical-point scaling motivated by high-temperature series expansions has recently been developed,
which aims to extend the scaling window well beyond the critical regime \cite{Ca}.
At a temperature $T$ sufficiently close to the critical temperature $T_c$,
divergent thermodynamical averages display the scaling behavior
$O(T)\approx A_O t^{-\rho}$ where $A_O$ is a constant amplitude, $\rho$ is
the critical index associated with the observable $O$ and
$\displaystyle{t=(T-T_c)/T_c}$
is the standard reduced temperature.
In order to improve the temperature range where scaling holds, other 
thermal scaling variables have been considered in the literature.
In particular the alternative reduced temperature 
\begin{equation}
\tau = \frac{T-T_c}{T} = \frac{\beta_c-\beta}{\beta_c} \,
\label{tau}
\end{equation}
is popular in analysis of series expansions. This variable 
is also natural in renormalization group analyses where the temperature 
variable usually appears through $\beta J$ ($J$ being a coupling constant 
between microscopic degrees of freedom), 
and thus the reduced variable is defined as in~(\ref{tau}).
The relationships between
the two reduced temperatures are
\begin{eqnarray}
 t  & = & \frac{\tau}{1-\tau} = \tau + \tau^2 + \tau^3 + \dots \,,
\label{tt}\\
 \tau & = & \frac{t}{1+t}     = t - t^2+t^3 - \dots \,,
\label{ttt}
\end{eqnarray}
such that in the vicinity of the critical point
\begin{equation}
 O(T) \approx A_O t^{-\rho} \quad \mbox{or} \quad O(T) \approx A_O \tau^{-\rho}\,.
\label{Ot}
\end{equation}
While both $t$ and $\tau$ vanish as the critical point is approached, they have very different 
high-temperature limits,
\begin{eqnarray}
 \lim_{T\rightarrow \infty}t & = & \infty \,,
\label{limitoft}
\\
 \lim_{T\rightarrow \infty}\tau & = & 1 \,.
\label{limitoftau}
\end{eqnarray}
In \cite{Ca}, a rationale was given explaining why the alternative reduced temperature $\tau$ in
(\ref{tau}) may be superior to the more traditional variable $t$ in scaling analyses,
at least in the  high-temperature regime. This rationale stems from  the following observations.
Firstly, suppose the constant $A_O$ in (\ref{Ot}) is promoted to a temperature-dependent 
amplitude as follows,
\begin{equation}
 O(T) \propto O^*(T)  \sim T^{\psi_O} (T-T_c)^{-\rho}
                \sim T^{\psi_O-\rho} \left({1-\frac{T_c}{T}}\right)^{-\rho} 
               \sim \beta^{\phi_O} \tau^{-\rho}\,,
\label{funamp}
\end{equation}
where $\phi_O = \rho - \psi_O$.
From (\ref{limitoftau}), 
 \begin{equation}
 \lim_{T\rightarrow \infty}{O^*(T)} \sim \beta^{\phi_O} \,,
\label{limoffc}
\end{equation}
and $\phi_O$ may be chosen so that $O^*(T)$ matches the high-temperature
series expansion (HTSE) for $O(T)$ in this limit.
Note that 
a similar approach cannot be implemented for scaling in $t$.
I.e., scaling in $\tau$, unlike  scaling in $t$, allows $O^*(T)$ to represent
the correct asymptotic behavior of $O(T)$ in the $T \rightarrow \infty$ limit
as well as close to criticality. Inspired by (\ref{funamp})
we now write the full expression for the observable $O(T)$ as
\begin{eqnarray}
 O(T) = A_O O^*(T)= A_O \beta^{\phi_O} \tau^{-\rho} + \dots\,,
\label{mess}
\end{eqnarray}
where the dots  represent higher-order additive corrections.

In \cite{Ca,CaBu}, Campbell et al. proposed and tested the extended
scaling scenario in the Ising model in two dimensions as well as the
Ising, Heisenberg and XY models in $d=3$ dimensions. These works
convincingly established the superiority of the method over the
conventional scheme between the lower and upper critical dimensions.
The inherent idealness of the  extended scaling approach to study systems
{\emph{below}} the lower critical dimension was demonstrated
in~\cite{CaKa}. In this work, we test extended scaling for the
$d$-dimensional Ising models with $d=5, 6, 7$ and 8, i.e.
{\emph{above}} the upper critical dimension $d_c=4$. 
Furthermore, while the $d \rightarrow \infty$ (mean-field) case was also investigated in \cite{CaBu}
(where extended scaling is exact  for all $T > T_c$),
here we include a detailed analysis of corrections in the extended scaling scheme.


In the next section we review the known results about the leading and
sub-leading scaling behavior of the Ising model above $d_c$. We then
define the extended scaling of this model, taking into account scaling corrections.
In Section~3 the MC approach is presented and in Section~4
the resulting data is discussed. Agreement with results from
HTSE confirms the efficacy of extended scaling, which is then used to
determine the critical parameters governing scaling in higher dimensions.
Since above $d=6$ dimensions, the dominant corrections to scaling are due
to analytic terms, this work provides a full account of leading and
confluent corrections in the susceptibility above the upper critical
dimension.  We conclude in Section~5.

\section{Extended scaling in high dimensions}
\setcounter{equation}{0}


Above the upper critical dimension $d_c=4$, the leading scaling behavior for the
magnetic susceptibility is given by mean field theory and is characterized
by the exponent $\gamma = 1$. The additive correction-to-scaling terms are
expected to be non-classical. The generic scaling form for the reduced
susceptibility for the Ising model is
\begin{equation}
  \chi (T)/\beta = \Gamma t^{-\gamma} \left({ 1 + b_1 t^{\theta} + b_2 t^{2\theta} + \dots
   + c_1 t + c_2 t^2 + \dots }\right) \,,
\label{generic1}
\end{equation}
where the dots indicate higher-order terms. In~(\ref{generic1}), the confluent
(or non-analytic) corrections involve the universal exponent $\theta$, while the remaining
correction terms are analytic.  (While in principle there could be several confluent 
correction exponents involved in (\ref{generic1}), it is sufficient in this work to
consider the single exponent $\theta$.) 
Analysis of the $\phi^4$ model above $d_c=4$ shows that the dominant 
critical behavior is determined by the Gaussian fixed point, leading to 
$\displaystyle{\chi \approx \langle{\phi^2}\rangle \approx t^{-1}}$ and $\displaystyle{\xi \approx t^{-1/2}}$. The 
confluent scaling corrections are due to the irrelevant $\phi^4$ term. The 
perturbation felt by the system inside the correlation volume is 
$\displaystyle{\langle{\phi^4}\rangle/\xi^d \approx \langle{\phi^2}\rangle^2/\xi^d}$ (since the 
average is taken over a Gaussian distribution), which thus behaves in the 
vicinity of the critical point as $\displaystyle{\chi^2/\xi^d\approx t^{-2+d/2}}$. As a 
consequence, the exponent $\theta$ is given by
\begin{equation}
 \theta = \frac{d-4}{2}
\,
\label{general}
\end{equation}
for $d>d_c=4$. Thus, for $d > 6$, where $\theta >1$, the confluent corrections
are expected to be overwhelmed by the analytic ones.  In six dimensions, the corrections
are expected to be modified by a logarithmic factor \cite{Gu81,Jo72}, and,
explicitly, 
\begin{eqnarray}
 \chi(T)/\beta  &= & \Gamma t^{-1} + B t^{-\frac{1}{2}} + C + D t^{\frac{1}{2}} + \dots\,,
\quad {\mbox{for $d=5$,}}
\label{d=5}\\
 \chi(T)/\beta & =  & \Gamma t^{-1} + B \ln{t} + C + D t\ln{t} + \dots\,,
\quad {\mbox{for $d=6$,}}
\label{d=6}\\
 \chi(T)/\beta & =  & \Gamma t^{-1} + C  + D t^{\frac{1}{2}} + \dots\,,
\quad {\mbox{for $d=7$,}}
\label{d=7}\\
 \chi(T)/\beta & =  & \Gamma t^{-1} + C  + D t + \dots\,,
\quad {\mbox{for $d=8$.}}
\label{d=8}\end{eqnarray}

Series analysis techniques have been used to verify the leading 
correction-to-scaling exponent $\theta$ in five and six dimensions 
and to determine some of the critical amplitudes \cite{Gu81}.
In particular, accepting the classical value $\gamma = 1$,
Guttmann measured $\theta=0.50(5)$, $\Gamma =1.311(9)$ and  $B= -0.48(3)$
in five dimensions as well as $\theta=0.98(5)$ and $\Gamma =1.168(8)$ 
in six dimensions \cite{Gu81}. It was not possible to determine the subdominant amplitudes,
nor to confirm the logarithmic term, in six dimensions using these techniques~\cite{Gu81}.

\begin{table}
\caption{Selection of previous estimates for the critical temperature $T_c$ for high-dimensional Ising models together with the
refined extended scaling estimates obtained in Sec.~4 of this paper. 
Here, FSS means a Monte Carlo approach using  finite-size scaling.}  
\begin {center}
\begin{tabular}{|l|l|l|l|l|l|l|}  \hline \hline
Reference method \& year                                     & $T_C$       & $T_C$    & $T_C$  &    $T_C$    \\
                                       &            $d=5$      & $d=6$    & $d=7$  &    $d=8$  \\
\hline

\cite{FiGa64}   $1/d$ expansion (1964)        &  8.7881\dots   & 10.8397\dots  & 12.8712\dots & 14.8923\dots \\
\cite{Gu81}   (critical expansion)  (1981)       &8.8162(5)& 10.8656(8)  &           & \\
\cite{MuHe93} HTSE (1993)          & 8.777(2)   &            &           & \\
\cite{GoAd93} HTSE (1993)                &            & 10.8348(4)  & 12.8690(4) & \\
\cite{LuBi99} FSS (1999)              &  8.77844(2)&   &   & \\
\cite{AkEr00} FSS (2000)       &            & 10.835(5)  &           & \\
\cite{AkEr01} FSS  (2001)     &            &            & 12.870(5)&\\
\cite{JoYo05} FSS (2005)               &  8.7785(5)&   &   & \\
\cite{MeDu06} FSS (2006)      &            &            &  12.870(5)         &14.893(3) \\
This paper     (2008)         &  8.7777(9)          &   10.8353(4)         &  12.8690(3)                &14.8933(8) \\
\hline \hline
\end{tabular}
\end{center}
\label{tab1}
\end{table}
The high-temperature series expansion (HTSE) for $d$-dimensional Ising models 
was given to fifteenth order in $\tanh(\beta)$ by Gofman et al. in \cite{GoAd93}, 
and compared with results of Monte Carlo simulations in six and seven dimensions. 
The HTSE in five dimensions was analyzed in a related study \cite{MuHe93}.
In contrast to five dimensions where the leading corrections are confluent, 
analytic corrections to scaling were favored in $d=6$ and $d=7$ 
dimensions \cite{GoAd93} using the HTSE approach. 
However, the logarithmic term in six dimensions was also not explicitly handled 
in the series expansion approach of \cite{GoAd93}. 
With no prior assumptions regarding exponent values, and ignoring the logarithm in 
six dimensions, the HTSE gave $\theta=0.45(10)$, $\theta = 1.0(3)$ 
and $\theta = 0.8 (2)$ in $d=5$, $d=6$ and $d=7$ dimensions, respectively
\cite{GoAd93,MuHe93}.
For convenience, a list of previous measurements of the critical temperatures
in high-dimensional Ising models is given in Table~\ref{tab1}.
\\

Here, we report on simulations of the Ising model in $d=5$, $6$, $7$ and $8$ dimensions
using the new extended scaling method recently developed by Campbell et al. \cite{Ca}.
It is claimed that this new method can extend the scaling regime well beyond that traditionally available \cite{Ca}. 
This claim has been verified in all models so far tested below their upper  critical dimensions \cite{CaBu}
and the method is particularly well suited to models which order at zero temoerature, i.e., below the {\emph{lower}} critical dimension \cite{CaKa}. 
A central theme of this paper is to test the applicability of the method 
for high dimensions, i.e., above the upper critical dimension, and
if it is proven more suitable than ``standard temperature scaling'', the 
method will be used to investigate corrections to scaling and amplitudes 
beyond the leading critical behaviour.

We consider the magnetic susceptibility $\chi(T)/\beta$ which is defined in the
paramagnetic phase as
\begin{equation}
  \chi(T)/\beta = \frac{1}{N} \sum_{i,j}{\langle{s_is_j}\rangle}\,,
\label{redchi}
\end{equation}
and numerically estimated by integration of the spin-spin correlation functions
over the lattice. With complete randomization in the high-temperature limit
($T \rightarrow \infty$), $\langle{s_is_j}\rangle = \delta_{ij}$ there, so that
$\chi(T)/\beta$ approaches unity.  Therefore the HTSE for the reduced susceptibility
has the form
\begin{equation}
 \chi(T)/\beta = a_0 + a_1\beta + a_2 \beta^2 + a_3 \beta^3 + \dots \,,
\label{HTE}
\end{equation}
where $a_0=1$ for all $d$.
Comparing this to (\ref{limoffc}), it is clear that $\phi_\chi=0$ 
for the reduced susceptibility.
This is the reason why the second equation in (\ref{Ot}) may represent the
susceptibility over an extended high-temperature range.
The limit (\ref{limitoft}), on the other hand, restricts the suitability of
the conventional variable $t$ for temperature scaling analysis in
the vicinity of the critical point.
Writing the five-dimensional critical expansion (\ref{d=5}) in terms of the 
extended scaling reduced temperature $\tau$ instead of $t$, one has
\begin{equation}
  \chi(T)/\beta = \Gamma \tau^{-1} + B\tau^{-\frac{1}{2}} + C(\tau)\,,
\label{fit5}
\end{equation}
where the higher corrections are contained in the function $C(\tau)$, which
goes to a constant $C$ in the high-temperature limit.
Since both $\tau$ and $\chi (T)$ approach unity there, one has
$C = 1-\Gamma -B$. Similar considerations in six and higher dimensions lead to 
\begin{equation}
  \chi(T)/\beta \sim \Gamma \tau^{-1} + B\ln{(\tau)} + C\,,
\label{fit6}
\end{equation}
 and
\begin{equation}
  \chi(T) /\beta \sim \Gamma \tau^{-1} +  C\,,
\label{fit7}
\end{equation}
respectively, where $C = 1-\Gamma $.

\section{The Monte Carlo algorithms}
\label{numerical}
\setcounter{equation}{0}

The (main) data presented in this work have been obtained using the so-called
worm algorithm introduced by Prokof'ev and Svistunov and which provides an
efficient Monte Carlo sampling of the dominant terms of the HTSE of spin-spin
correlation functions~\cite{Prokofev01,Wolff08}. The data resulting from this
approach are labelled MCHTSE (for Monte Carlo high-temperature series
expansion) throughout this paper.
In the case of the Ising model, this HTSE is
\begin{equation}
\langle s_is_j\rangle
={1\over{\cal Z}}\sum_{\{s\}}s_is_j
e^{\beta \sum_{(k,l)} s_ks_l}
={1\over{\cal Z}}(\cosh \beta )^{dN}
\sum_{\{s\}}s_is_j
\prod_{(k,l)} \big(1+s_ks_l
\tanh\beta \big)
\,.
\end{equation}
Here, $s_i$ is the spin at site $i$ of the $d$-dimensional lattice,
${\cal Z}$ is the partition function, the summation is taken over
configurations and  $(k,l)$ denotes a pair of neighboring sites on the lattice. 
Since $\sum_{s_i=\pm 1}{s_i}^{2p}=2$ while $\sum_{s_i=\pm 1}{s_i}^{2p+1}=0$ 
 for integer $p$,
the only graphs that contribute to the expansion are paths joining sites
$i$ and $j$ (which we  call sources) and closed loops.
These graphs can be sampled stochastically using the following rules: one of the
sources, say that initially at site $i$, is moved to one of the neighboring sites,
say $i'$, and a bond is added between $i$ and $i'$ if none was hitherto present,
or the bond is deleted if already present. If the two sources are on the same site,
i.e. $i=j$, they are both moved to a new  randomly chosen site. These three
moves are chosen using a Metropolis prescription: the probability to add a bond
is ${\rm min}(1,\tanh\beta J)$, that to delete a bond is ${\rm min}(1,1/\tanh\beta J)$
while the probability to move both sources can be set freely. This algorithm
is known to be slightly more efficient than Swendsen-Wang cluster
algorithm~\cite{Sokal07}. To accelerate the dynamics, the state of the bonds
starting from the same site is stored as a single bit. Furthermore we
implemented a continuous-time version of the worm algorithm. At each Monte Carlo
step, the probability $\omega_\alpha$ of all possible moves is calculated.
The time $\tau$ that the system will stay in the same state is evaluated as~\cite{Bortz75}
\begin{equation} 
\tau=1+{\rm Int}\left({\ln\eta\over 
\ln(1-\sum_\alpha\omega_\omega)}\right)\,, 
\end{equation}
where $\eta$ is a random variable uniformly distributed over $]0;1]$. The time
is increased by $\tau$ and one of the moves $\alpha$ is chosen with
probability $\omega_\alpha$ and applied to the system.
\\

\begin{figure}[t]
\vspace{9cm}
\includegraphics{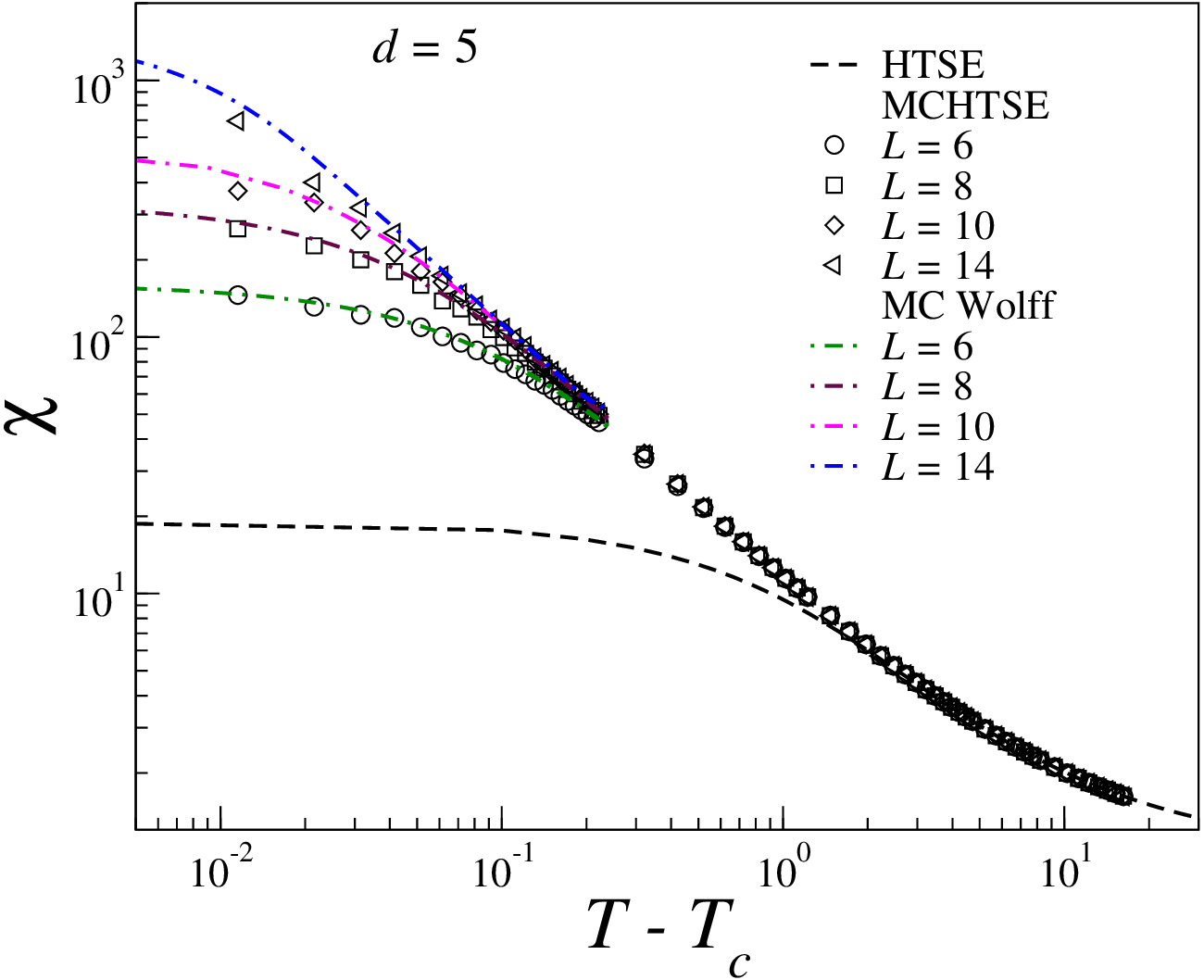}
\caption[a]{MCHTSE estimates for the susceptibility of the five-dimensional Ising
model compared with conventional Monte Carlo estimates (dot-dashed 
curves, colored online) using the Wolff algorithm together with histogram reweighting
and with the HTSE of \cite{GoAd93} (dashed line, black online).}
\label{fig:1d5}
\end{figure}

As the temperature approaches the critical temperature, the worm algorithm
generates graphs contributing to higher and higher orders of the HTSE.
The number of these graphs grows exponentially fast which may
cause critical slowing-down of the convergence of the averages.
To check the results, additional data have been computed using the
standard Wolff algorithm~\cite{Wolff89} whose critical slowing-down is much
better understood. For a given lattice size, the
simulations have been performed at a temperature close to $T_c$, and
data for nearby temperatures were obtained by the so-called histogram
reweighting method \cite{Ferr&Swen88&89}.
The reliability of the MCHTSE data is demonstrated in Fig.~\ref{fig:1d5},
where estimates for the susceptibility from this method are compared with
those from the HTSE of \cite{GoAd93} and conventional Monte Carlo data
for lattice sizes $L=6$--$14$ in five dimensions. Clearly, the MCHTSE data
accurately follow the HTSE curve for large $T$ and are comparable to
conventional MC estimates closer to $T_c$, where finite-size effects
become important. Furthermore, the MCHTSE for $L \ge 10$ is
independent of $L$ down to $T \approx T_c +0.1 \approx 8.9$.

\section{Extended scaling analysis of MC data}
\setcounter{equation}{0}

Using the numerical approach presented in Sec.~\ref{numerical}, we
computed the magnetic susceptibility $\chi$ in Ising models above $d_c$.
Extended scaling is tested in Figs.~\ref{fig:1d5}--\ref{fig:3d8}
and then employed to determine the leading corrections for the scaling behavior
of the susceptibility. In particular, we firstly determine $\theta$, $\Gamma $,
$B$ and $C$ in five dimensions. The agreement between our measurements there and
those of \cite{Gu81} establishes confidence in the extended scaling approach.
The fifteenth-order HTSE for $d$-dimensional Ising models of Gofman et al. in
\cite{GoAd93} is indeed expected to be accurate at sufficiently high temperature,
while the critical expansion of Guttmann \cite{Gu81} should be reliable close to 
criticality. Extended Scaling is then used to measure $\theta$, $\Gamma $, $B$ and $C$
in six dimensions as well as $\theta$, $\Gamma $ and $C$ in seven and eight dimensions. 
The scaling forms (\ref{d=5}), (\ref{d=6}), (\ref{d=7}) and (\ref{d=8}) are confirmed
(including for the first time the logarithmic correction in the leading
correction-to-scaling term in six dimensions) and hence the general formula (\ref{general})
is supported. These forms are then used to determine refined estimates for $T_c$ 
in $d=5$ to $d=8$ dimensions. The amplitude of the logarithm in six dimensions turns out
to be small, explaining the difficulties in verifying it numerically almost 
thirty years ago~\cite{Gu81}.

\subsection{Five-dimensional case}
\begin{figure}[t]
\vspace{9cm}
\includegraphics{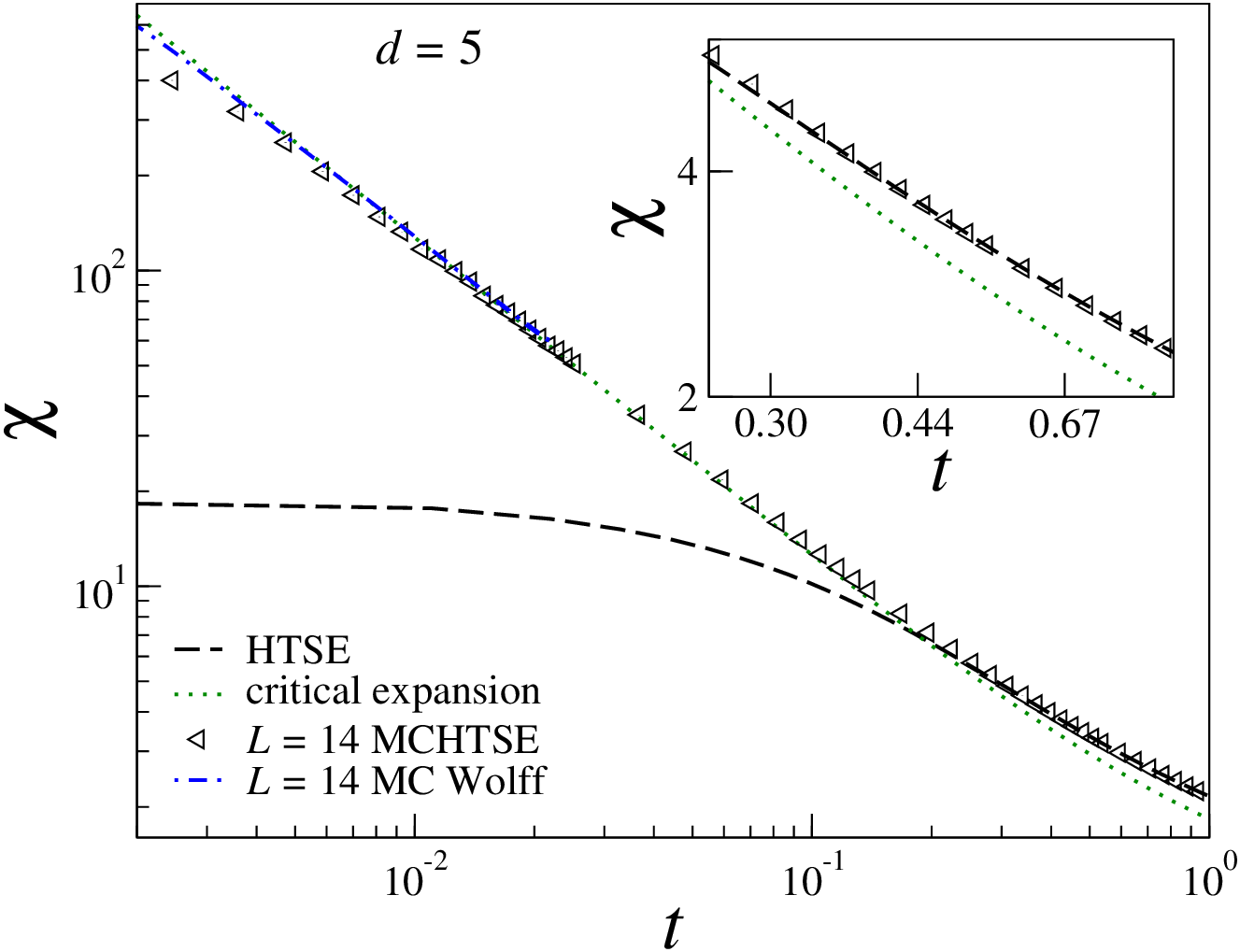}
\caption[a]{MCHTSE estimates for the susceptibility of the five-dimensional 
Ising
model compared with conventional Monte Carlo estimates 
(dot-dashed curve, blue online), the HTSE of \cite{GoAd93}
(dashed line, black) and the critical expansion 
of \cite{Gu81} (dotted line, green online).
The inset illustrates how the critical expansion becomes
inaccurate at high temperature where the MCHTSE data cross over to 
the HTSE curve.}
\label{fig:2d5}
\end{figure}

In Fig.~\ref{fig:2d5}, the $d=5$, 
$L=14$ MCHTSE data are compared with the HTSE \cite{GoAd93}, the critical 
expansion \cite{Gu81}  and conventional $L=14$ MC data  on a double-logarithmic scale using the 
standard reduced temperature $t$.
The MCHTSE data follow the critical curve for
$t{\rm{\raisebox{-.75ex}{ {\small \shortstack{$>$ \\ $\sim$}} }}} 0.05$
to
$t{\rm{\raisebox{-.75ex}{ {\small \shortstack{$<$ \\ $\sim$}} }}} 0.2$,
where they cross over to the 
HTSE curve. 
The insert in Fig.~\ref{fig:2d5} is a blow-up of this region,
and clearly illustrates the deviation of  the critical expansion  from 
the HTSE and the MCHTSE data.

To extract the parameters governing the scaling in five
dimensions, the following extended scaling  analysis was applied to the 
MCHTSE data. 
From (\ref{d=5}), the first three terms characterizing the critical behavior 
of the susceptibility are expected to be of the form
\begin{equation}
  \chi(T)/\beta \sim \Gamma \tau^{-\gamma} + B\tau^{\vartheta} + C\,,
\label{fit55}
\end{equation}
where $\vartheta = -\gamma + \theta$ from (\ref{generic1}), 
$\tau$ is given in (\ref{tau}) and  extended scaling gives $C=1-\Gamma -B$.
Firstly, for $L=14$, 
a six-parameter fit of the MCHTSE data to (\ref{fit55}) between
$T=8.9$ and $T=25$ gives an estimate for $T_c$ of $8.7743(95)$,
which compares well with the best estimate in the literature 
($T_c = 8.77844(2)$  \cite{LuBi99}).
Refined extended scaling estimates for the critical temperatures at each value of $d$ 
are given later in this section and listed in Tables~\ref{tab1} and~\ref{tabresults}. 
Accepting this more accurate estimate for $d=5$, a fit to the remaining
five parameters yields an estimate for $\gamma$ of $1.0(2)$.
Accepting the mean field value $\gamma = 1$ and fitting to the remaining four parameters gives 
$\Gamma =1.28(2)$, $B=-0.44(19)$, so that $1-\Gamma -B= 0.16(21)$, which is compatible with
$C=0.17(21)$. Accepting that $C=1-\Gamma -B$ (i.e., using extended scaling), 
a three-parameter fit to the amplitudes $\Gamma $ and $B$, as well as to the correction
exponent gives $\vartheta= -0.47(8)$.
Finally, accepting that $\vartheta$ is, in fact $-0.5$, as in (\ref{d=5}),
a two-parameter fit to the amplitudes of the leading and first correction terms gives
$\Gamma  = 1.291(3)$ and $B=-0.310(8)$.
The goodness of fit for each of these measurements (and for 
each fit reported below) is monitored through the chi-squared per degree of freedom and is observed to be
reasonable in each case.
These results, together with the higher dimensional ones, are summarized in Table~\ref{tabresults},
where the refined (see below) extended scaling estimates for the critical temperature are also given.

The same analysis applied to the $L=16$ MCHTSE data yields very similar results
and, in particular, $\Gamma  = 1.291(3)$,       $B=-0.311(8)$.
For comparison, Guttmann's series analysis yielded
$\Gamma  = 1.311(9)$ and $B=-0.480(30)$ \cite{Gu81}.

\begin{figure}[t]
\vspace{9cm}
\includegraphics{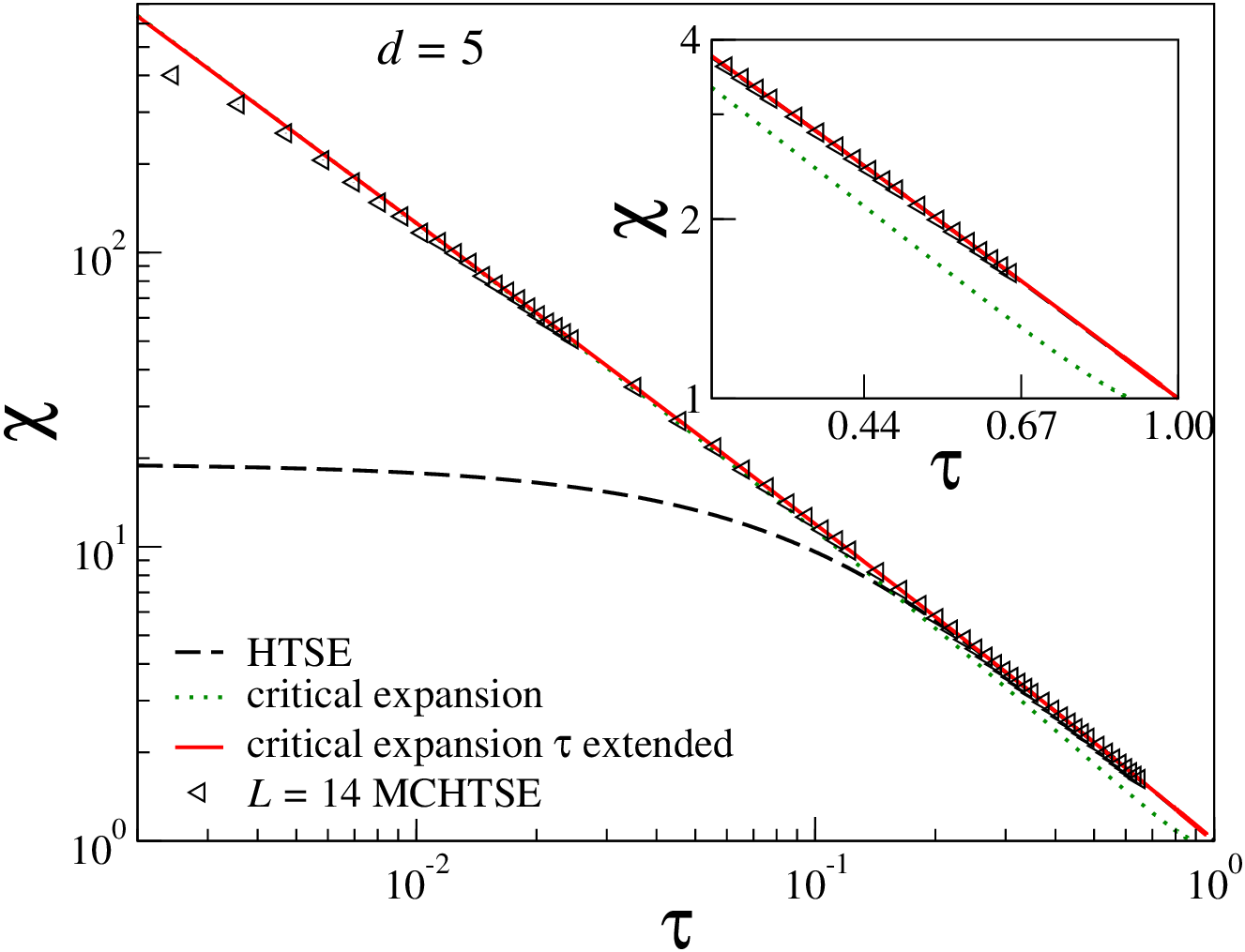}
\caption[a]{Extended scaling for the susceptibility of the $d=5$ Ising model.
The extended scaling curve (solid, red online) 
coincides with the critical curve 
(dotted, green online) \cite{Gu81}
close to $\tau=0$ ($T=T_c$) and with the HTSE (dashed, black) 
\cite{GoAd93} close to $\tau=1$ 
($T \rightarrow \infty$) as well as with the MCHTSE estimates in between.}
\label{fig:3d5}
\end{figure}

This extended scaling  
(i.e., (\ref{fit5}) with $C(\tau)=C=1-\Gamma -B$ and $\Gamma  = 1.291(3)$ and   $B=-0.311(8)$)
is depicted in Fig.~\ref{fig:3d5} for the $d=5$ case. 
This line coincides with the HTSE \cite{GoAd93}
 for high temperature ($\tau \rightarrow 1$)
and with the critical expansion \cite{Gu81} close to $T_c$ ($\tau=0$). 
Between these extremes, it successfully follows the MCHTSE data.
It is clear from Fig.~\ref{fig:3d5} that 
the critical reduced-temperature line of  \cite{Gu81}
deviates significantly from the data away from $T_c$,
a circumstance  typically ascribed to correction terms.
Similarly the  HTSE curve deviates from the data as the temperature is reduced.
On the other hand, the  reduced temperature curve nicely follows the critical expansion 
for small $\tau$ and the data
to very large $T$. 
Thus the  extended scaling method proposed in \cite{Ca} is indeed seen to be superior to 
the conventional approach, also in high dimensions.

In summary, the extended scaling analysis of the MCHTSE data 
yields similar but improved results to that of  \cite{Gu81}
for the $d=5$ Ising susceptibility.
Therefore, confidence has been established in the method, which can now be applied
to higher dimensions to determine critical parameters which were 
unobtainable previously.

\subsection{Six-dimensional case}
\begin{figure}[t]
\vspace{9cm}
\includegraphics{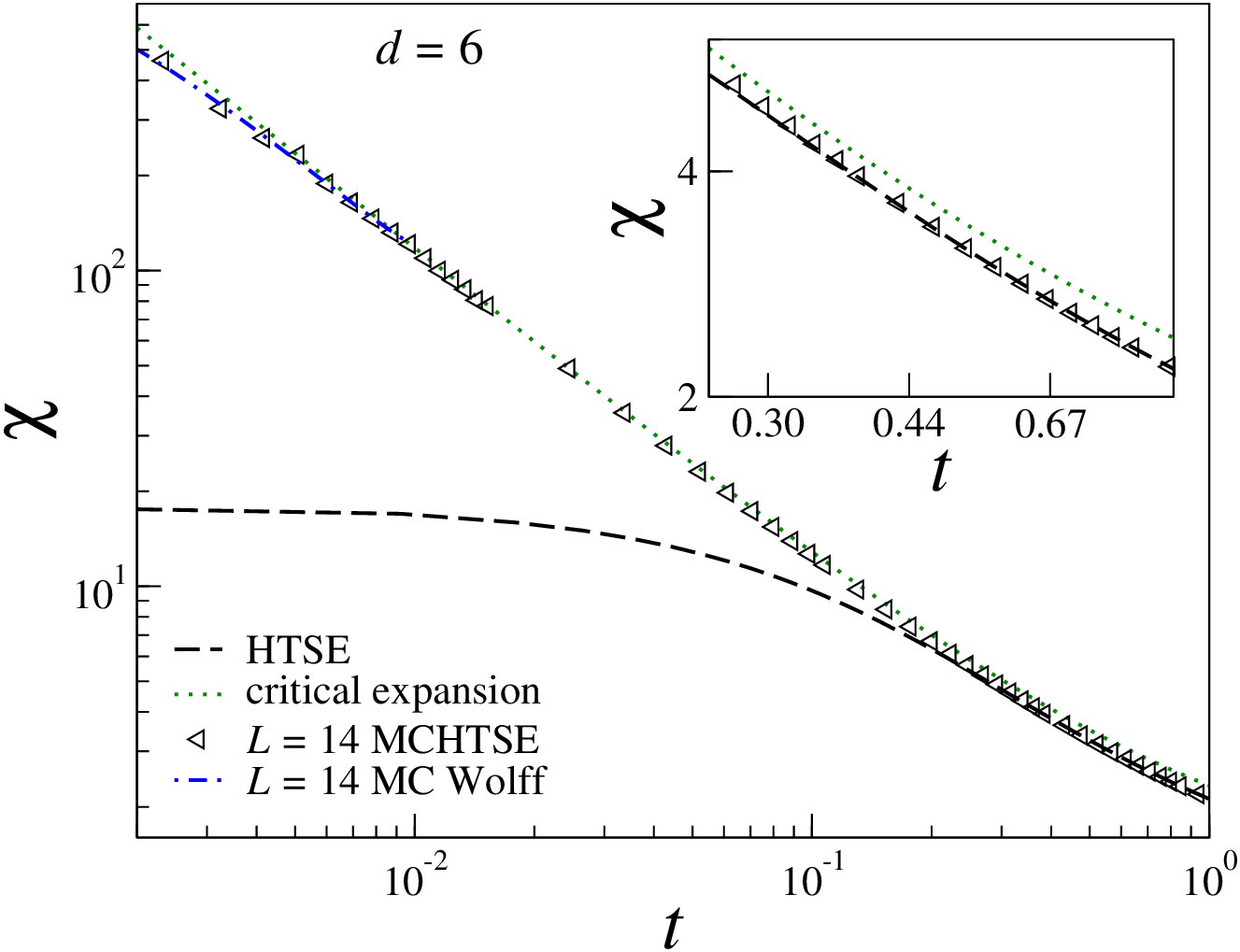}
\caption[a]{MCHTSE estimates for the susceptibility of the 
six-dimensional Ising
model compared with estimates coming from conventional Monte Carlo 
(dot-dashed, blue online), the HTSE of \cite{GoAd93} 
(dashed, black)
and the critical expansion of \cite{Gu81} (dotted, green online).
The inset illustrates the deviation of the  critical expansion from
the HTSE at high temperature.}
\label{fig:2d6}
\end{figure}

In Fig.~\ref{fig:2d6}, the six-dimensional
$L=14$ MCHTSE data are compared with the HTSE
and the critical 
expansion  as well as  conventional (reweighted) $L=14$ MC data.
Similar to the five-dimensional case, 
the MCHTSE data follow the critical curve close to $t=0$
and switches to the HTSE curve for larger $t$. 
The insert of Fig.~\ref{fig:2d6} illustrates the deviation of  the critical curve
from the HTSE and the MCHTSE data.

Although the leading correction term in six dimensions is expected to 
involve a logarithm after (\ref{d=6}), we initially fit to the form
(\ref{fit55}) so as not to a priori bias in favor of the logarithmic structure.
Using the MCHTSE data for $L=14$,
and fitting to $55$ data points from  $T=10.88$ to $T=30$,
a six-parameter fit to the form (\ref{fit55}) gives $T_c=10.8318(24)$, close to the value $T_c=10.8348(4)$
 measured in \cite{GoAd93}.
Following the same procedure as in the five-dimensional case, and accepting this
 value, subsequent fits yield 
$\gamma = 0.95(25)$ and $\vartheta = -0.01(40)$.
Since the latter value is close to zero (and because the fitting process becomes
more unstable as the number of parameters is reduced), we 
conclude that a logarithmic term may indeed be present at the leading-correction
level.

We therefore perform a four-parameter fit to the form (\ref{fit6})
which gives $\gamma = 0.994(3)$.
Accepting the mean field value $\gamma = 1$, 
a three-parameter fit gives  $1-\Gamma = -0.1607(24)$ and $C= -0.1605(18)$.
Accepting that $C=1-\Gamma $ (extended scaling) then 
yields $\Gamma  = 1.1606(17)$ and $B = 0.0571(27)$.
The value for $\Gamma $ compares well with Guttmann's estimate $1.168(8)$ \cite{Gu81}.
The logarithmic term was not addressed in previous analysis \cite{Gu81,GoAd93}.

\begin{figure}[t]
\vspace{9cm}
\includegraphics{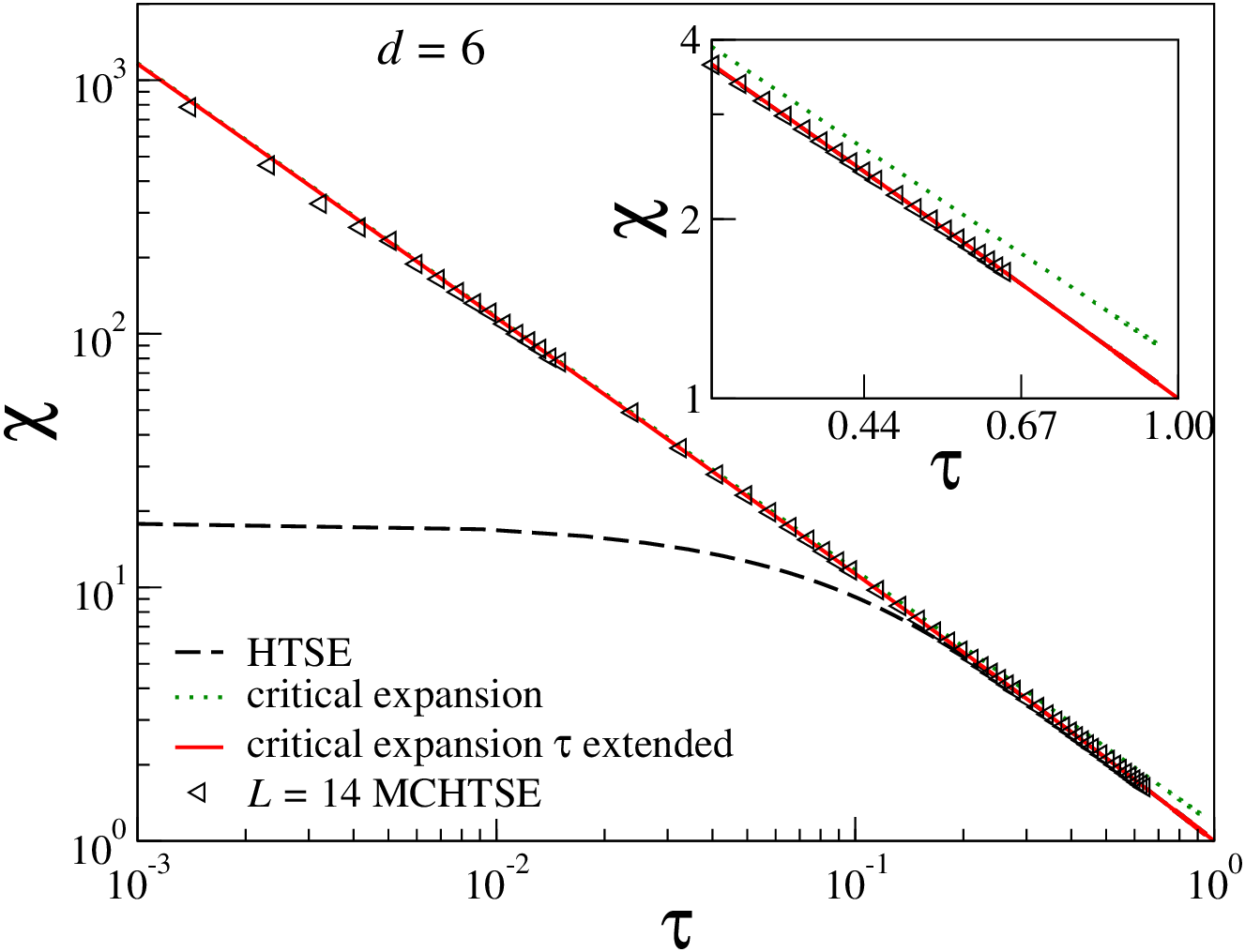}
\caption[a]{Extended scaling for the susceptibility of the Ising model
in $d=6$ dimensions.
The  extended scaling curve (solid, red online) 
traces the critical curve 
of \cite{Gu81} (dotted, green online) 
close to $\tau=0$  and the  HTSE  \cite{GoAd93} (dashed) close to $\tau=1$ 
and as with the MCHTSE estimates in between.}
\label{fig:3d6}
\end{figure}
Extended scaling  (i.e., (\ref{fit6}) with  $\Gamma  = 1.1606(17)$, $B = 0.0571(27)$ and $C=1-\Gamma $)
is depicted  in Fig.~\ref{fig:3d6} for the $d=6$ case. 
This  line coincides with the HTSE \cite{GoAd93}
 for high temperature ($\tau \rightarrow 1$)
and with the critical expansion \cite{Gu81} close to $T_c$ ($\tau=0$)
and traces the MCHTSE data in between.

\subsection{Seven-dimensional case}
\begin{figure}[t]
\vspace{9cm}
\includegraphics{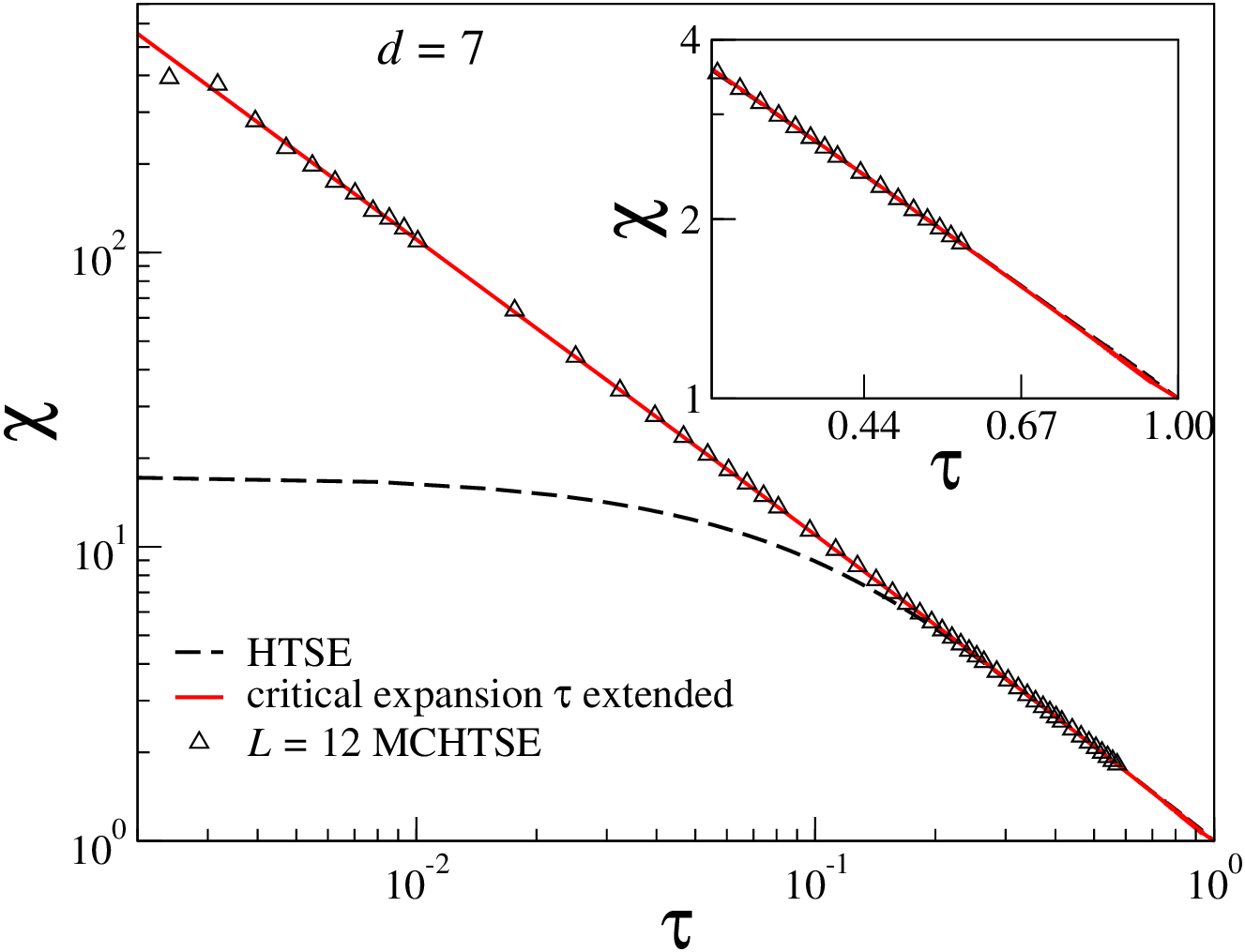}
\caption[a]{Extended scaling (solid curve, red online) for the susceptibility of the Ising model
in $d=7$ dimensions with the HTSE of \cite{GoAd93} (dashed curve, black online) and the MCHTSE data.}
\label{fig:3d7}
\end{figure}
For $d=7$ dimensions, we simulated at three different lattice sizes ($L=10$, $12$ and $14$)
to again determine the temperature range over which the MCHTSE data are $L$ independent
($T=12.89$--$30$).
Fitting the $L=12$ data 
to all four parameters in (\ref{fit7}) yields $T_c=12.8665(5)$, close to 
the value $12.8690(4)$   
(obtained in  \cite{GoAd93} using the HTSE), which we now accept
(pending our refined extended scaling estimate, which we give below).
A three-parameter fit then yields $\gamma = 1.000(2)$,
and accepting the mean field value $\gamma = 1$, we find
$\Gamma =1.1086(9)$ and $C=-0.1215(20)$.
Accepting that $C=1-\Gamma $ (extended scaling) finally gives $\Gamma = 1.1008(5)$.
The extended scaling plot for seven dimensions is given in
Fig.~\ref{fig:3d7}.

\subsection{Eight-dimensional case}
\begin{figure}[t]
\vspace{9cm}
\includegraphics{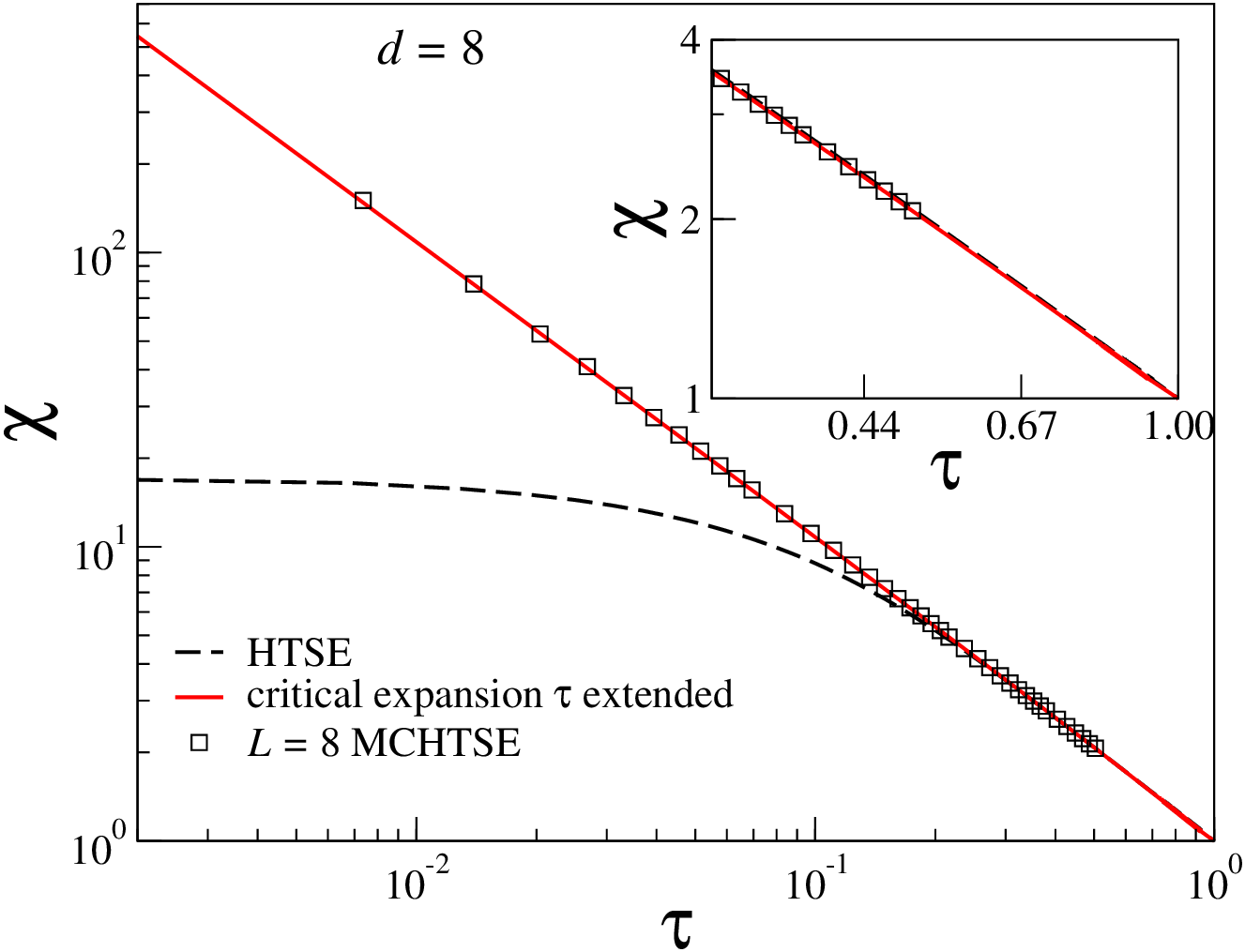}
\caption[a]{Extended scaling curve (solid, red online) for the Ising susceptibility 
in $d=8$ dimensions with the HTSE of \cite{GoAd93} (dashed curve, black online) and the MCHTSE data.}
\label{fig:3d8}
\end{figure}

A similar analysis in eight dimensions (with $L=8$)
 gives the successive estimates 
$T_c=14.8893(20)$  (to be compared with the value $T_c=14.893(3)$ from 
finite-size scaling  \cite{MeDu06}),
$\gamma = 0.998(2)$, $\Gamma =1.0935(45)$ and $C=-0.0950(26)$.
Extended scaling ($C=1-\Gamma $) then yields $\Gamma =1.0836(5)$.
The extended scaling plot for $d=8$ is depicted in Fig.~\ref{fig:3d8},
where it too is compared with the HTSE \cite{GoAd93} and the critical expansion \cite{Gu81}.
These results, with refined estimates for the critical temperature, are summarized in Table~\ref{tabresults}.
~\begin{table}[h]
\caption{Summary of the numerical results obtained in this paper for dimensions $d=5$ to $d=8$.
The leading exponent is compatible with the mean field result $\gamma=1$. 
In $d=6$ dimensions, the leading correction is logarithmic. Analytic
corrections dominate confluent exponents beyond $d=6$. }
\begin {center}
\begin{tabular}{|l|l|l|l|l|}  \hline \hline
 $d$   &$T_c$           &$\gamma$ &$\vartheta  $& Amplitudes    \\
\hline
 $5$    &$ ~8.7777(9)$  &$1.0(2)$      &$-0.47(8)$ & $\Gamma  = 1.291(3)$, $B=-0.310(8)$    \\
     &   &      &  &                                                            $C=0.019(11)$   \\
 $6$    &$10.8353(4)$ &$0.994(3)$  & $0$ [logarithm]   &    $\Gamma  = 1.1606(17)$, $B = 0.0571(27)$    \\
          &                      &                     &             &   $C=-0.1606(17)$   \\
 $7$    &$12.8690(3)$ &$1.000(2)$ &                 &   $\Gamma = 1.1008(5)$, $C=-0.1008(5)$  \\
 $8$    &$14.8933(8)$ &$0.998(2)$  &                 &  $\Gamma =1.0836(5)$, $C=-0.0836(5)$    \\
\hline \hline
\end{tabular}
\end{center}
\label{tabresults}
\end{table}

Having established that $\gamma=1$ in each case, and the expected scaling forms (\ref{d=5}) to (\ref{d=8})
indeed hold, we return to the estimates of $T_c$ in each dimension.
Refitting to these forms for the critical temperature (as well as the amplitudes)
yields the more refined estimates
$T_c=  8.7777(9)$, 
$T_c=10.8353(4)$, 
$T_c=12.8690(3) $, 
$T_c=14.8933(8)$, 
for $d=5$, $d=6$, $d=7$ and $d=8$ respectively.
These values should be compared with previous estimates in the literature, which are listed in Table~\ref {tab1}.

\begin{figure}[t]
\vspace{8cm}
\includegraphics{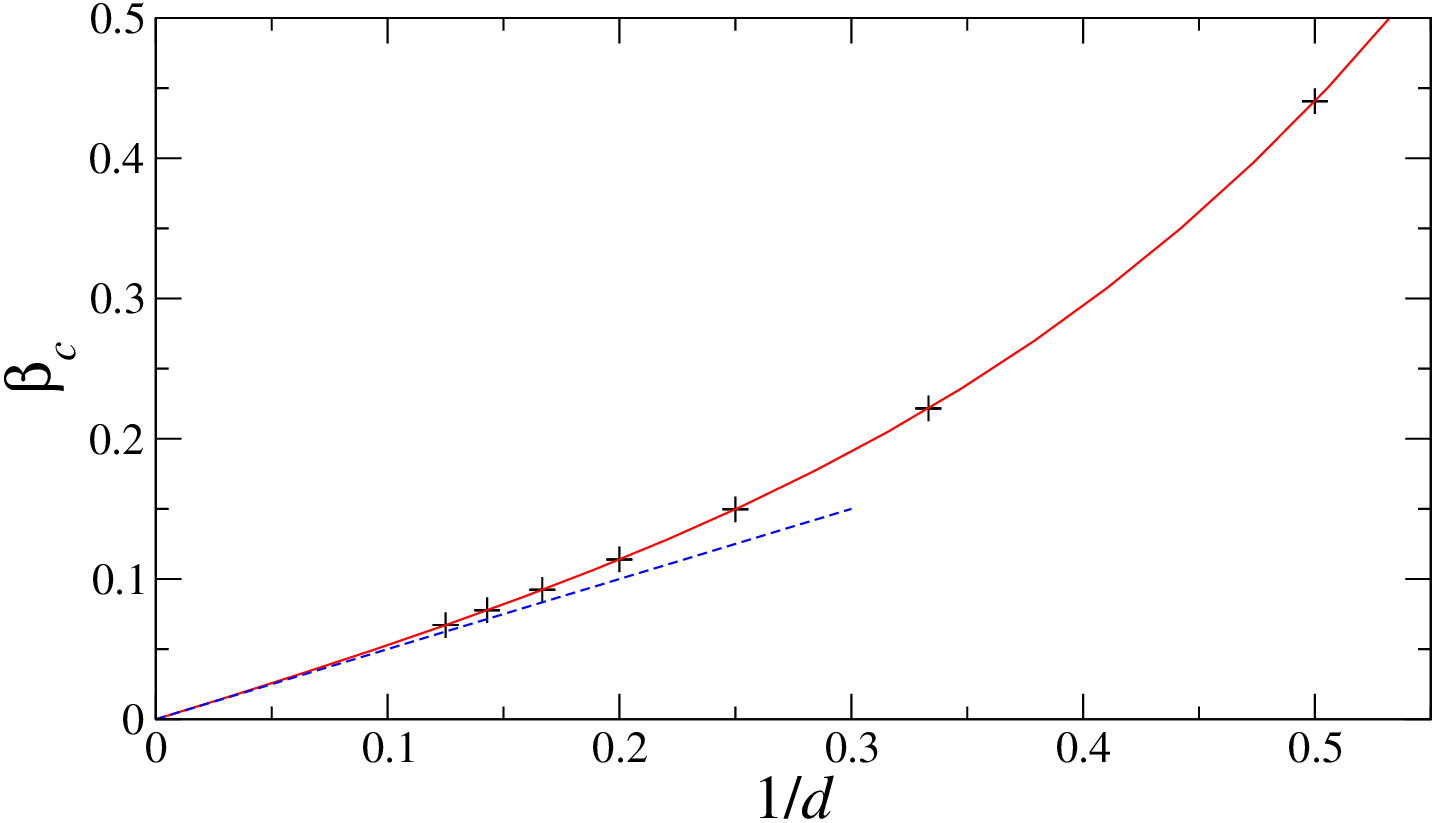}
\vspace{-0cm}
\caption[a]{
The approach of the critical temperature (solid curve, red online)
to the classical mean field  value (dashed line, blue online)
with increasing dimensionality.}
\label{MF1}
\end{figure}

With the Boltzmann constant and the coupling strength having both been set to unity,
for a regular lattice of the type considered here the mean field expressions for the Ising critical temperature 
and amplitude are $T_c = 2d$ and $\Gamma = 1$. 
The results for $T_c$ and $\Gamma $, summarized in Table~\ref{tabresults}, illustrate this approach to
these classical values as the dimensionality is increased. 
The approach to mean field is also illustrated in Fig.~\ref{MF1}, using our
estimates for $T_c$ for $d=5$ to $d=8$ together with the exact result for $d=2$ \cite{KrWa41} and best estimates from the 
literature for $d=3$ and $d=4$ \cite{DeBl03,KeLa93}. 
These data are fitted to polynomials of increasing degree in $1/d$ until the goodness of fit becomes reasonable.
The solid curve  represents a fit to the polynomial form $\displaystyle{\beta_c = b_1/d + b_2/d^2 + b_3/d^3 + b_3/d^4}$,
with  $b_1 = 0.5044(3)$, $b_2 = 0.1935(30)$, $b_3=0.3540(101)$ and $b_4=1.5335(104)$. 
The dashed line  represents the mean field approximation $T_c = 2d$.

\section{Conclusions}

We have presented estimates for the magnetic susceptibility in the 
paramagnetic phase of high-dimensional Ising models,
obtained using the Prokof'ev-Svistunov worm algorithm
which enables the stochastic generation of the HTSE by Monte
Carlo methods. The resulting data are checked against 
the critical expansion \cite{Gu81}, the HTSE \cite{GoAd93} and 
the results of a standard Monte Carlo approach using the Wolff 
algorithm \cite{Wolff89}.
When applied to this data, the new extended scaling approach of
\cite{Ca} is demonstrated to work well in $d=5$ dimensions, where
known results for the estimates of critical parameters \cite{Gu81} 
are recovered and improved. Having thus
established confidence in the extended scaling approach above the upper critical dimension, 
it is then applied to the data obtained in $d=6$, $d=7$ and $d=8$ dimensions
to deliver estimates for critical parameters (especially those 
governing confluent corrections) there.
In particular, the logarithmic correction in six dimensions 
(unobtainable in previous analyses \cite{Gu81,GoAd93}) is clearly
verified. It is also observed how the infinite dimensional 
($d\rightarrow \infty$) limit leads to the mean field theory.
In this way, the general formula (\ref{general}) is supported 
for the Ising model and a full account of the leading and confluent
corrections to scaling in the odd sector above the upper critical 
dimension  is given.

~\\
\noindent
{\bf{Acknowledgements:}}
The laboratoire de Physique des Mat\'eriaux is Unit\'e Mixte de Recherche
CNRS number 7556. We wish to thank Paolo Butera and Ian Campbell for
e-mail correspondences.

\bigskip
%

\end{document}